# Probing terahertz surface plasmon waves in graphene structures

Oleg Mitrofanov,[1,2,*] Wenlong Yu,[3] Robert J. Thompson,[1] Yuxuan Jiang,[3] Igal Brener,[2,4] Wei Pan,[4] Claire Berger,[3,5] Walter A. de Heer,[3] and Zhigang Jiang[3,**]

[1] University College London, Electronic and Electrical Engineering, London, WC1E 7JE UK

[2] Center for Integrated Nanotechnologies, Sandia National Laboratories, Albuquerque, NM 87185, USA

[3] School of Physics, Georgia Institute of Technology, Atlanta, GA 30332, USA

[4] Sandia National Laboratories, Albuquerque, NM 87185, USA

[5] CNRS/Institut Néel, BP166, 38042 Grenoble, France

**Abstract:** Epitaxial graphene mesas and ribbons are investigated using terahertz (THz) near-field microscopy to probe surface plasmon excitation and THz transmission properties on the sub-wavelength scale. The THz near-field images show variation of graphene properties on a scale smaller than the wavelength, and excitation of THz surface waves occurring at graphene edges, similar to that observed at metallic edges. The Fresnel reflection at the substrate SiC/air interface is also found to be altered by the presence of graphene ribbon arrays, leading to either reduced or enhanced transmission of the THz wave depending on the wave polarization and the ribbon width.

Graphene and its nanostructures have recently emerged as a promising platform for next-generation nano-plasmonic devices[1-6]. It has been shown experimentally that patterned graphene structures, such as graphene ribbon arrays, are capable of hosting surface plasmons with long lifetime and a high degree of optical field confinement, owing to the exceptionally low ohmic loss of the material[1, 2, 7-9]. The frequency of the plasmon mode is continuously tuneable within the THz spectral range by varying the array dimension and the carrier density in graphene. The plasmon modes can potentially be used in a range of device applications from tuneable THz filters and THz sensors to graphene-based information processing and communications.

Due to the localized nature of THz surface plasmon waves in graphene, their direct experimental detection needs to be performed in a region near the surface. Near-field microscopy provides this capability and can enable investigations of plasmon effects in graphene [4, 5, 10]. Recently, THz near-field microscopy with an integrated sub-wavelength aperture probe has proven to be a sensitive tool in exploring THz surface waves on metallic surfaces[11, 12]. In application to graphene, this method potentially allows for probing the local transmission properties (with a spatial resolution of 5-10μm[13]), as well as the processes of plasmon mode excitation by the incident THz radiation and the processes of plasmon wave propagation along the surface. It is worth noting that far-field THz spectroscopy and imaging of graphene has already been conducted by several groups[14,15,16]. These techniques enable non-contact characterization of the THz conductivity of graphene and provide direct information about the doping level and sample quality. Spatial resolution of the far-field methods is however limited by diffraction.

In this Letter, we report on near-field THz imaging of epitaxial graphene mesas and ribbon arrays and provide a side-by-side comparison with a noble metal (Au) film. We demonstrate that the local THz transmission properties of graphene structures are modified by the presence of confined plasmon mode (in ribbons) and that the propagating surface plasmon waves are excited at the edges of graphene.

Multilayer epitaxial graphene samples (3.4nm and 7.1nm thick) grown on the C-face of SiC are patterned into mesa and ribbon array structures via electron-beam lithography in a large writing field, followed by oxygen plasma etching and vacuum annealing[9]. The samples are illuminated by a beam of broadband (0.5-2.5THz) THz pulses generated through optical rectification in a ZnTe crystal[12] and delivered via a hollow cylindrical THz waveguide[17] (1.6mm in diameter), as shown in Fig. 1(a). The waveguide provides a narrow angular spectrum distribution of the incident wave[17]. It also preserves the incident THz field distribution with respect to the imaged sample. In this experimental arrangement, the near-field images can be interpreted as the electric field distribution formed near the graphene structures illuminated by a ~0.5mm diameter THz beam (intensity FWHM), fixed with respect to the sample.

It is important to note that in the near-field region there are three possible contributions to the detected THz electric field: (1) the transmitted wave, (2) the confined plasmon mode (in ribbons) with an in-plane wave vector ($q \neq 0$) corresponding to the ribbon array period and (3) the surface plasmon wave that can be excited at the pattern edges ($q \approx 0$). These contributions are coherent and therefore their interference must be taken into account. To separate the interference effects from the intrinsic transmission properties, the detected signal is analyzed in the time domain rather than in the frequency domain: the fields due to the surface plasmon and the ribbon plasmon mode are expected to exhibit a temporal delay with respect to the transmitted wave. In this work, near-field images are obtained with a 10μm aperture probe scanning over the sample at a distance of less than 10μm. The corresponding spatial resolution is therefore ~10μm and any variation of graphene properties on a smaller scale cannot be resolved.[11]

First we consider the effect of graphene ribbon arrays on the transmitted THz field using an array consisting of 7.1nm thick and 200nm wide ribbons separated by 200nm. The ribbons, 400μm in length, form square-shaped ribbon arrays (400μm × 400μm), as illustrated schematically in Fig. 1(b). The ribbons are oriented parallel to the incident THz pulse polarization (along the *x*-axis). Figures 1(c,d) show a near-field THz image of the sample and an image of the beam profile (after sample is removed) respectively, with the electric field amplitude taken at the time delay corresponding to the pulse center. The detected field is primarily dominated by the transmitted wave contribution.

Quantitative analysis of Fig. 1(c) reveals that the amplitude of the transmitted electric field is decreased by ~20% in the areas covered by graphene, making the ribbon squares darker in the near-field image. The Fourier transform of the transmitted and incident pulses shows that the transmission coefficient is frequency independent throughout the spectral range of our measurement (0.5-2.5THz). This result is consistent with our recent work on the same sample[9], where the plasmon mode of top graphene layers is found to be at $\omega_{pl} \approx$ 4.6THz. $\omega_{pl}$ is expected to be at an even higher frequency for the first few layers at the SiC-graphene interface due to the higher carrier concentration. The confined plasmon modes in graphene ribbons[1,8,9] or domain structures[3] therefore are not responsible for the observed transmission properties.

The local transmission varies slightly (±5%) across the graphene areas and the variation appears to be more pronounced near the square edges/corners. We note that the observed THz transmission variation is partially due to the non-uniform thickness of multilayer epitaxial graphene[18], found in optical microscopy (see also Fig. 2(a)) and atomic force microscopy measurements (not shown).

To investigate further the origin of non-uniform field detected over the surface of graphene, near-field transmission properties are mapped for a 3.4nm thick control sample consisting of 200μm × 200μm graphene mesas. Figure 2 presents an optical image of the mesa and the instantaneous THz near-field images at two moments within the duration of THz pulse (Fig. 2(b)). Here the highest contrast is observed in the image corresponding to the amplitude of the incident field approaching zero: $E_{inc}(t) = 0$, e.g., at $t = 2.23$ps. Panels (c) and (d) in Fig. 2 compare the near-field images measured with two different scan steps, and repeatable electric field patterns are observed across the mesa. The patterns are partially correlated with visible variations (owing to thickness variation) in the optical image of Fig. 2(a). A similar pattern is also observed in Figs. 2(e,f), recorded at $t = 1.25$ps, at the negative peak of the first cycle of the THz pulse.

The detected field in Fig. 2 contains two contributions: the transmitted wave and surface waves excited at the mesa edges. The evidence of surface wave excitation at the mesa edges can be extracted from the THz space-time maps and evaluated in comparison with this effect observed on the noble metal surface. Figure 3(a) shows a THz image of a 200nm thick Au reference square (left) and a 3.4nm thick graphene mesa (right). The thickness of the Au square is large enough to block the THz wave completely, therefore any THz field detected in the Au area is due to the surface wave excited at the edge[12].

To probe the surface wave excitation, space-time scans are taken along the horizontal and vertical directions across both graphene and Au structures. In the case of Au, typical surface wave signatures are observed: the space-time wavefronts become tilted in the Au region in the scan parallel to the polarization of the incident field, whereas a phase shift is present for the orthogonal scan[12, 13, 19]. These maps indicate that the surface waves are excited from the opposite sides of the Au square and they are traveling along the metallic surface. The surface waves arrive at the center of the square with a time delay of ~0.3ps, giving rise to the phase shift in the vertical space-time scan of Fig. 3(c).

In order to observe surface waves on graphene mesa, the transmitted THz wave ($E_{tr}$),

which gives the largest contribution to the detected signal, must be subtracted from the detected field ($E_{det}$). As the transmission coefficient exhibits no significant frequency dependence, $E_{tr}(t)$ can be assumed to be a replica of the incident field ($E_{inc}$) scaled by a constant $a$, $E_{tr}(t) = a \times E_{inc}(t)$, where $a \approx 0.8$ (for the graphene mesa sample) and $E_{inc}(t)$ is an average THz pulse waveform measured over an area of the sample where no graphene is present. Space-time maps of $E(x, t) = E_{det}(x, t) - a \times E_{inc}(t)$ and $E(y, t) = E_{det}(y, t) - a \times E_{inc}(t)$ are plotted in Figs. 3(d,e). Similar to the 200nm thick Au square, both the *xt*- and *yt*-maps show signatures of a surface wave excited from the opposite side of the graphene mesa. The surface waves propagate along the graphene surface and arrive at the mesa center with a temporal delay. The similarity of the space-time maps for the metallic square and the graphene mesa indicate that the surface wave excitation at the mesa edges is the most probable mechanism. We note that the incident field is determined by the waveguide geometry and it has no *k*-vector components that can couple to surface plasmons directly. Therefore the only possible mechanism is the surface wave excitation at a surface discontinuity, which is the mesa edge in our sample.

Having considered the non-resonant effects observed on graphene mesas and ribbon arrays with small period, we now discuss the effect of ribbon arrays for which the plasmon mode frequency of the top graphene layers is expected to be $\omega_{pl} \sim 2$THz, given that $\omega_{pl} \propto (W)^{1/2}$. The ribbons are $W = 1\mu$m in width and 200μm in length, and the spacing between the adjacent ribbons is 1.2μm. Two principal polarizations of the incident THz pulse, with electric field parallel or perpendicular to the ribbon direction, are examined (Figs. 4(a,b)). In the first case, the ribbon periodicity is expected to exhibit no significant effect on the transmission properties, whereas it directly affects frequency of the confined plasmon mode in the latter case. Indeed, the observed THz transmission of the 1μm wide graphene ribbon arrays varies noticeably for the two orthogonal polarizations. This behavior is markedly different from that of 200nm wide ribbon arrays with 200nm spacing (Fig. 1(c)) for which we find no noticeable difference in transmission for both polarizations in the spectral range from 0.5-2.5THz.

Figures 4(c-f) summarize the THz space-time maps for two line scans with the ribbon orientation parallel and perpendicular to the incident THz field polarization. The change in the transmission coefficient is evident in Fig. 4(e,f), where $E_{det} - a\,E_{inc}$ is plotted. For the ribbons parallel to the field, the *xt*-maps exhibit reduction of the detected field due to the presence of graphene ribbons on the right-hand side of the map. This reduction is ~27+/-5% compared to the SiC/air interface and it is slightly stronger than the reduction observed for the graphene mesa (~20+/-5%) in Fig. 3. The difference is likely to be caused by additional absorption or scattering on ribbon edges. When the ribbons are oriented perpendicular to the field, however, the transmitted field is found to be noticeably stronger compared to the field detected over the blank SiC substrate. By adjusting the value of *a*, we determine that the amplitude transmission for the ribbon array in this case is increased by ~50% compared to that of the SiC/air interface.

The reduced/enhanced transmission shows the effect of the ribbons on properties of the SiC/graphene/air interface. The Fresnel reflection coefficient at the SiC/air interface is relatively high due to the large refractive index of SiC ($n \approx 3.0$). ~25% of the incident wave (intensity) is reflected at the SiC/air interface. The ribbons oriented perpendicular to the polarization appear to provide better impedance matching for the interface and lead to a higher transmission coefficient. For this orientation, the confined plasmon mode is expected to form in the ribbons array. The plasmon mode may results in the observed effect of enhanced transmission. A similar effect of enhanced transmission due to plasmons in metals has been observed previously for sub-wavelength hole arrays[20] and slit arrays in metallic films.[21] It must be noted however that THz wave coupling into the near-field probe may be different for the surface plasmon mode and for the transverse incident wave. Further investigations are needed for quantifying this effect.

Frequency of the surface plasmon mode can be tuned by chemical doping, $\omega_{pl} \propto (n_{el})^{1/4}$, where $n_{el}$ is the carrier density. Exposing graphene ribbons to moisture or solvent (blown dry by $N_2$) is expected to induce p-type carriers[22] and to result in a higher plasmon mode frequency (presumably outside of our measurement frequency range). Figures 2(c-f) show the THz images of the same sample as in Fig. 4 after chemical doping, where the ribbon array, located in the top half of the imaged area (Fig. 2(a)), exhibits no contrast compared to SiC surface. This observation implies that the enhanced THz transmission of undoped graphene ribbons (treated by vacuum annealing) might be due to the coupling of the incident THz wave to the plasmon mode.

In summary, we have performed a THz near-field microscopy study of epitaxial graphene mesas and ribbon arrays and observed surface waves excited at the edges of graphene structures. We find that the THz transmission through graphene ribbon arrays on SiC can be either reduced or enhanced, depending on the orientation of the ribbon with respect to the polarization of the THz wave. The enhanced transmission is observed in the near field zone for ribbons with the expected plasmon mode frequency of top graphene layers within the spectral range of the incident THz pulses (0.5-2.5THz). Chemical doping of graphene and reducing the ribbon periodicity, which lead to an increase of the plasmon mode frequency, show reduced THz transmission. These intriguing properties of graphene hold promise for new applications in THz spectroscopy, sensing, imaging, and communications. The application of the THz near-field microscopy technique also opens the possibility of non-invasive probing of local THz properties of graphene with sub-wavelength spatial resolution (~10 µm in this work) and for investigations of surface plasmon waves in graphene structures.

This work is supported by the Royal Society [grant number UF080745] and performed at UCL and at the Center for Integrated Nanotechnologies, an Office of Science User Facility operated for the U.S. Department of Energy (DOE) Office of Science by SNL [contract DE-AC04-94AL85000]. Epitaxial graphene growth and device fabrication are carried out at GaTech, supported by the NSF [DMR-0820382] and the DOE Office of Basic Energy Sciences (BES) through a contract with SNL. The work at SNL is supported by LDRD.

**References:**

*E-mail address: o.mitrofanov@ucl.ac.uk

**E-mail address: zhigang.jiang@physics.gatech.edu

Figure Captions

Fig. 1. (a) Schematic diagram of the near-field system for THz microscopy of graphene samples. Schematic diagram (b) and the corresponding THz near-field image (c) of graphene ribbon arrays arranged in 400μm × 400μm squares. The spacing between the squares is 100μm. The electric field amplitude in (c) is normalized to the field in the center of the THz beam measured without the sample (d).

Fig. 2. (a) Optical image of a 3.4nm thick epitaxial graphene mesa (bottom) and a 1μm wide graphene ribbon array (top). The ribbons are oriented along the vertical direction. (b) Time-domain waveform of the incident THz pulse measured by positioning the near-field probe over the SiC substrate. Arrows indicate two moments at which the near-field images are recorded. (c-f) THz images of the mesa area in (a), showing the instantaneous electric field at $t$ = 2.23ps (c,d) and at $t$ = 1.25ps (e,f). The left/right panels of the same $t$ represent consecutive measurements with step size 6.25μm/12.5μm.

Fig. 3. (a) THz image of a 3.4nm thick 200μm × 200μm epitaxial graphene mesa (right) and a 200nm thick 207μm × 207μm Au square (left). The incident THz pulse is polarized along the horizontal direction. (b-e) Space-time maps measured along the horizontal (b,d) and vertical (c,e) directions through the square centers: for the Au square, maps (b) and (c) show the detected electric field, $E_{det}$ (normalized to the peak value); for the graphene mesa, maps (d) and (e) show the difference of the detected field and the scaled incident field, $(E_{det} - a \times E_{inc}) / E_{inc}$, (where $a$ = 0.8) to emphasize the effect of surface plasmon waves.

Fig. 4. Optical images of 1μm wide graphene ribbon arrays (graphene thickness is 3.4nm, the ribbons are 200μm in length, and the spacing between the adjacent ribbons is 1.2μm) oriented parallel (a) or perpendicular (b) to the incident electric field vector. Panels (c) and (d) show the space-time maps measured across the pattern edge, while panels (e) and (f) emphasize the difference in electric field, $(E_{det} - a \times E_{inc}) / E_{inc}$, with $a$ = 0.73 for (e) and $a$ = 1 for (f), demonstrating the reduced (e) and enhanced (f) transmission through the ribbons with respect to the SiC substrate.

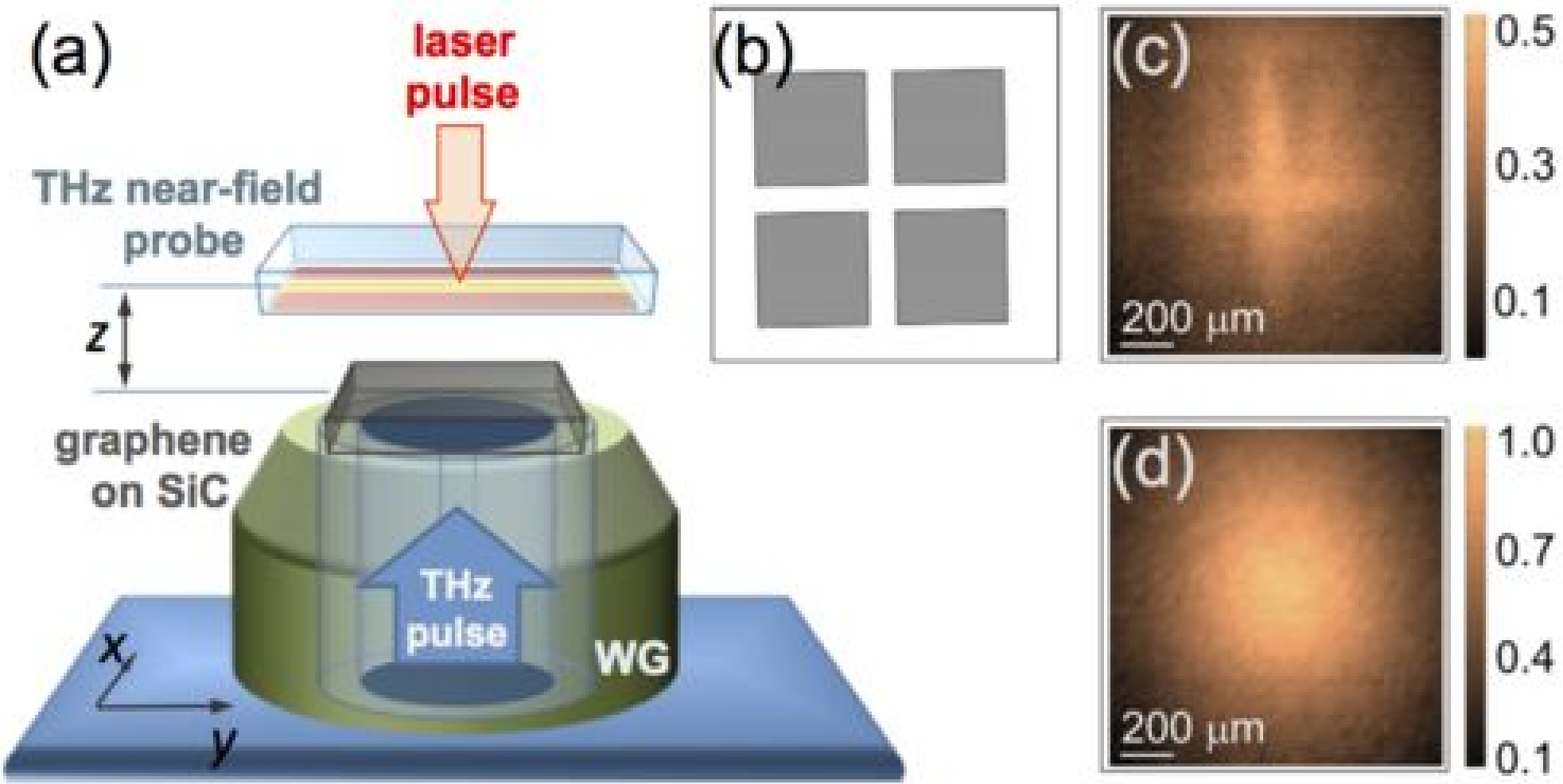

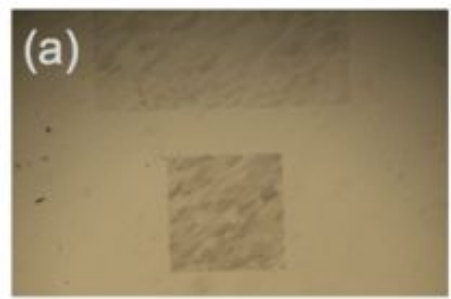
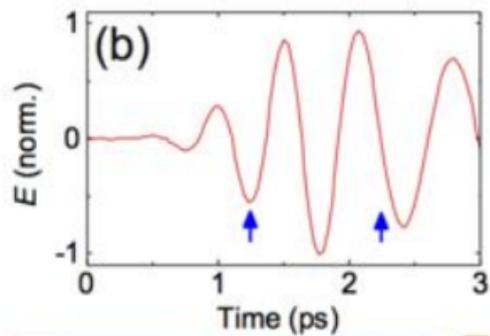
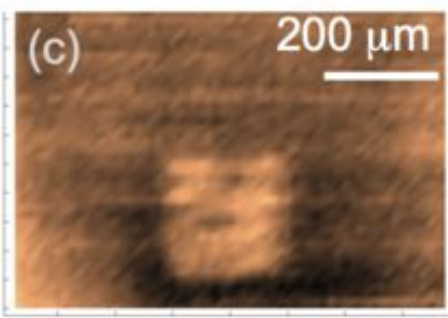
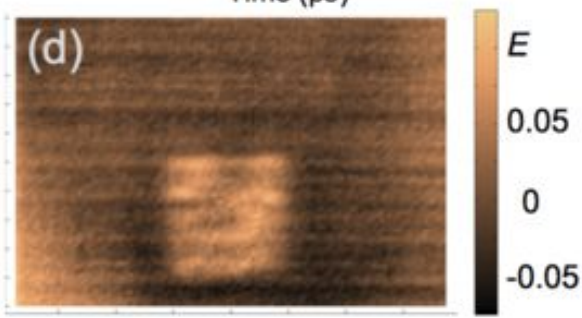
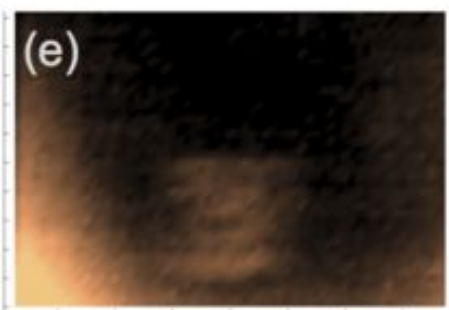
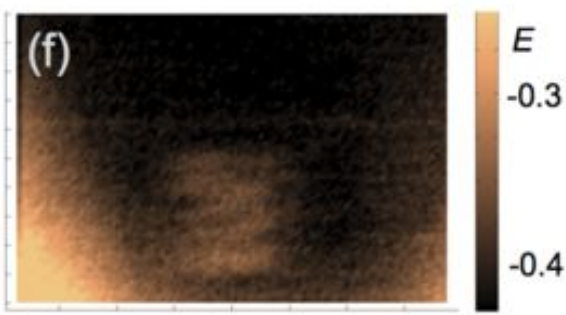

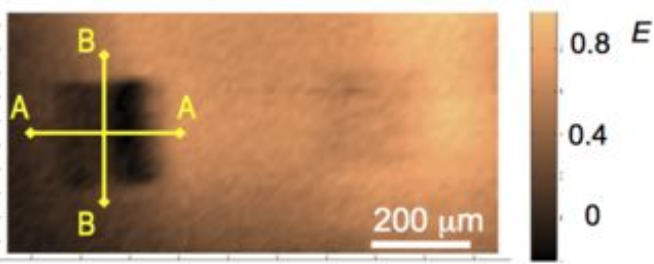
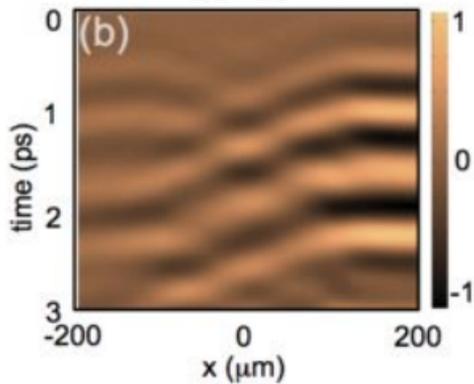
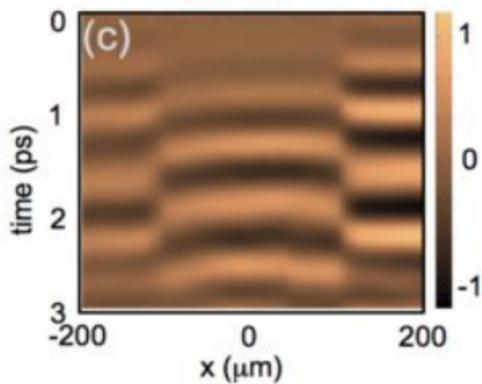
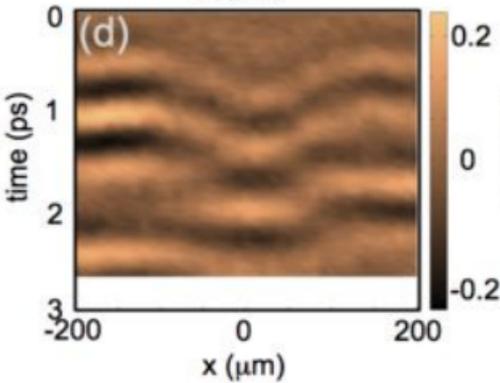
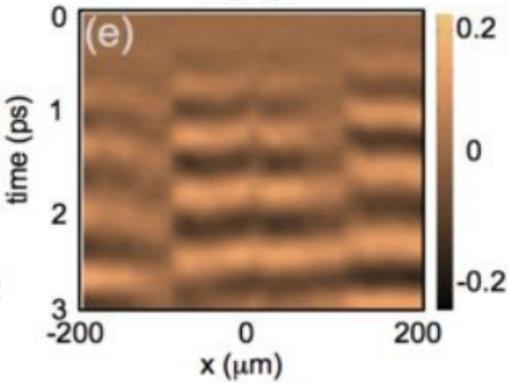

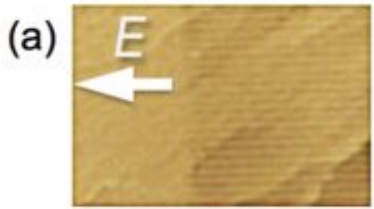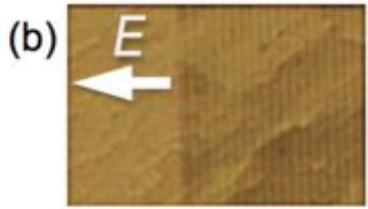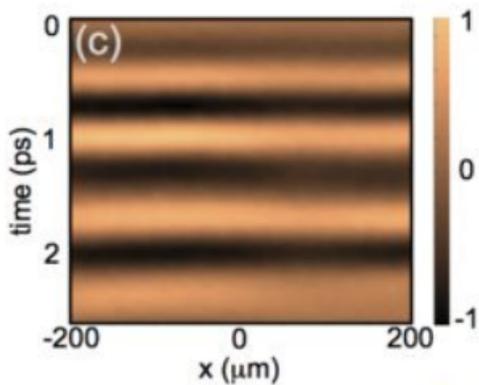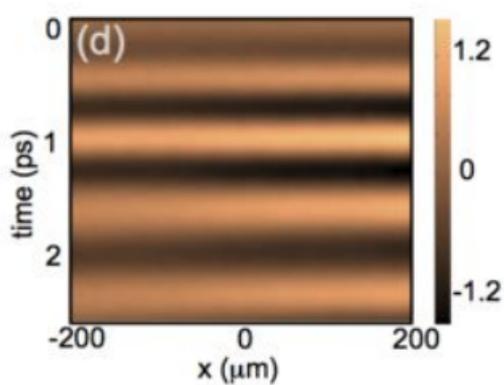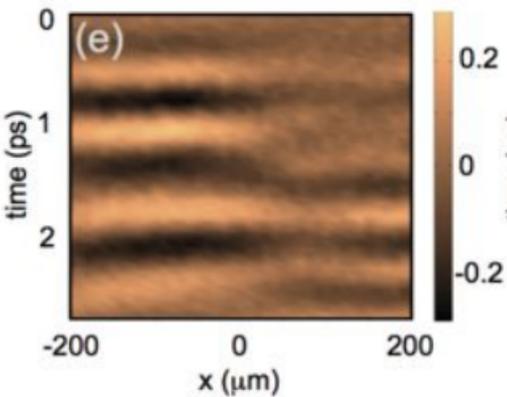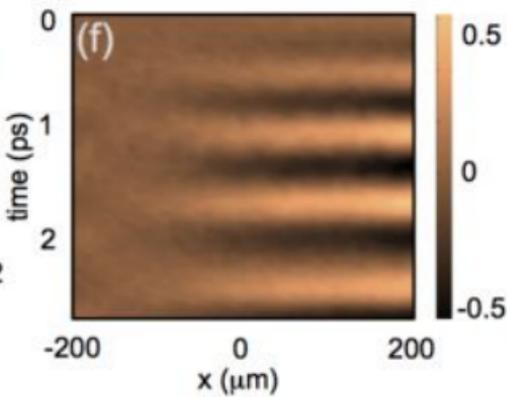